\documentclass[final]{Rinton-P9x6}
\usepackage{epsfig,mathrsfs,psfrag,amsmath,subfigure}
\newcommand*{\etal}{\emph{et~al.}}

\title{O(N) linear sigma model beyond the Hartree approximation
  at finite temperature
  \footnote{\lowercase{talk given by} S. M\lowercase{ichalski}.}
}
\author{J\"urgen Baacke}
\author{Stefan Michalski}
\address{Institut f\"ur Physik, Universit\"at Dortmund \\
  Otto-Hahn-Str. 4, D--44221 Dortmund \\
  eMail: \texttt{stefan.michalski@udo.edu}}

\begin{document}
\maketitle

\abstracts{
We  study  the $O(N)$ linear sigma model with spontaneous
symmetry breaking at finite temperature in the framework of the
two-particle point-irreducible (2PPI) effective action.
We go beyond the Hartree approximation by including the two-loop 
contribution, i.e., the sunset diagram.
A phase transition of second order is found, whereas it is of first order
in the one-loop Hartree approximation.
Furthermore, we show the temperature-dependence of the variational mass
parameters and comment on their relation to the physical sigma and pion 
masses.
}


\section{Introduction}
Nowadays it is common belief that the $O(N)$ linear sigma model 
has a second-order phase transition. Nevertheless, there is still
recent progress on this subject as people develop new approximation
methods of  the effective potential at finite temperature.

Inspired by earlier investigations of the non-equilibrium properties
of the model in the Hartree approximation\cite{Baacke:2001zt} 
-- where we found
striking similarities to the situation in thermal equilibrium --, we 
wanted to know if a higher approximation is able to model the correct order
of the phase transition.

Among recent publications there are, e.g., those by Bordag and 
Skalozub\cite{Bordag} and by Phat \etal\cite{Phat:2003eh} 
who used the 2PI CJT formalism
with a ``manually localized'' Dyson--Schwinger equation and found
a first-order phase transition even for higher loop-order.
In contrast to that there is the 2PPI formalism developed by Verschelde 
\etal\cite{Verschelde:bs} which
is able to mimic the correct order of the phase transition
for plain $\Phi^4$ theory\cite{Smet:2001un}.

We extend their 2-loop 2PPI analysis
of plain $\Phi^4$ theory to a full $O(N)$ model.
For more details we refer the reader to a more comprehensive 
publication\cite{Baacke:2002pi} and references therein.


\section{2PPI Effective Potential}
The classical action of the $O(N)$ linear sigma model in 3+1 dimensions
is given by
\begin{equation}
\mathcal{S}[\Phi] = \int d^4x\ \mathscr{L}[\Phi]
= \int d^4x \left[ \frac{1}{2} \partial_{\mu} \vec\Phi \cdot
   \partial^{\mu} \vec\Phi
   - \frac{\lambda}{4} \left(\vec\Phi\cdot\vec\Phi - v^2 \right)^2
   \right]\ , 
\end{equation}
where $\vec\Phi = \left( \Phi_1,\Phi_2,\dots,\Phi_N \right)$
is an $O(N)$ vector.
Using the formalism of the 2PPI effective 
action\cite{Baacke:2001zt,Verschelde:bs,Smet:2001un}
we eventually find an expression for the temperature-dependent
effective action (temperature times Helmholtz's free energy)
\begin{equation}
\label{eq:gamma2PPI}
\Gamma^\mathrm{2PPI}_\mathrm{eff}[\phi_i,\mathcal{M}^2_{ij}] = 
     \mathcal{S}[\phi] 
     + \Gamma_\mathrm{class}^\mathrm{2PPI}[\mathcal{M}^2_{ij}]
     + \Gamma_q^\mathrm{2PPI}[\phi_i,\mathcal{M}^2_{ij}] \ .
\end{equation}
An intrinsic property of this formalism is that the
propagator, which is used to evaluate the loop diagrams of
the quantum corrections, is \emph{naturally} a local one
\[ G_{ij}^{-1}(p^2) = p^2\ \delta_{ij} + \mathcal{M}^2_{ij}\ . \]

The variation of the effective action with respect to the 
mass matrix $\mathcal{M}^2_{ij}$ 
yields the gap equation
\begin{equation}
  \label{eq:gapij}
  \mathcal{M}^2_{ij} = \lambda (\phi^2-v^2)\ \delta_{ij}
  + 2\lambda\ \phi_i \phi_j 
  + \hbar\lambda\ \left( \sum_\ell \Delta_{\ell\ell}\ \delta_{ij} 
    + 2 \Delta_{ij}   \right)\ ,
\end{equation}
where $\hbar\Delta_{ij}$ is the local self-energy or the connected part of
the expectation value of the local composite operator $\Phi^2$. 
In Eq.~\eqref{eq:gapij} it enters as an explicit function of $\phi_i$
and $\mathcal{M}^2_{ij}$ given by the derivative
\begin{equation}
\Delta_{ij}(\phi_i,\mathcal{M}^2_{ij}) = 2\ 
     \frac{\delta \Gamma_q^\mathrm{2PPI}[\phi_i,\mathcal{M}^2_{ij}]}
     {\delta \mathcal{M}^2_{ij}}  \ .
\end{equation}

\begin{figure}[htbp]
  \centering
  \includegraphics[scale=0.2]{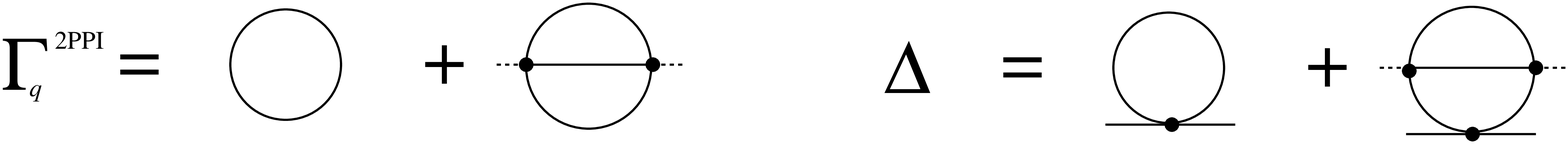}  
  \caption{Graphical representation of the effective action and 
    the self-energy in the two-loop approximation.}
  \label{fig:gamma2loop}
\end{figure}

Using an $O(N)$-invariant decomposition 
of the (variational) 
mass matrix $\mathcal{M}^2_{ij}$
  \[ \mathcal{M}^2_{ij} = \frac{\phi_i\phi_j}{\vec\phi^2} \mathcal{M}_\sigma^2
     + \left( \delta_{ij} - \frac{\phi_i\phi_j}{\vec\phi^2} \right) 
     \mathcal{M}_\pi^2\]
and restricting $\vec\phi = (\phi,0,\dots,0)$
we find the following gap equations 
\begin{subequations}
\label{eq:gapsigmapi}
\begin{eqnarray}
  \mathcal{M}^2_\sigma &=& 
  \lambda \left[ 3 \phi^2 - v^2 
    + 3\hbar\Delta_\sigma + (N-1)\ \hbar\Delta_\pi \right]
  \\
  \mathcal{M}^2_\pi &=& \lambda \left[ \phi^2 - v^2 + \hbar\Delta_\sigma +
    (N+1)\ \hbar\Delta_\pi \right] \ .
\end{eqnarray}
\end{subequations}
The 2-loop approximation is used here so that the 
the quantum corrections of the effective potential consist of
the 1-loop ``log det'' term plus the sunset graph.
The graphs contributing to the effective action and to the local 
self-energy are displayed in Fig.~\ref{fig:gamma2loop}.


\section{2-Loop Effective Potential at Finite Temperature}
\begin{figure}[htbp]
  \centering
  \subfigure[Hartree approximation for temperatures
  $T/v=1.46, 1.47,1.48, 1.49$.]{
    \psfrag{V}{$V_\mathrm{eff}(\phi)$}
    \psfrag{f}{$\phi$}
    \includegraphics[scale=0.2]{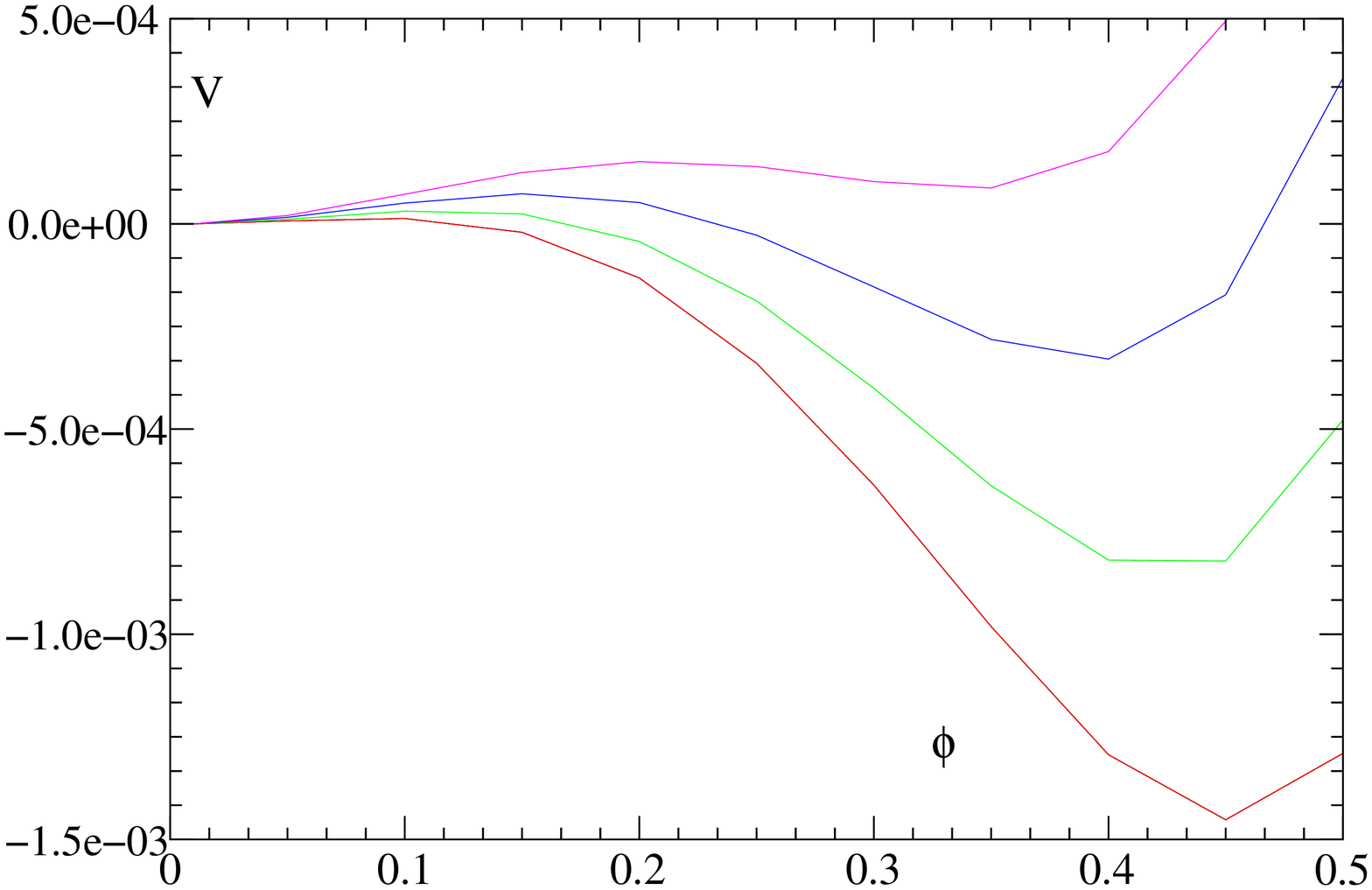}
  }
  \subfigure[2-loop approximation for temperatures
  $T/v=1.62, 1.66, 1.69, 1.70, 1.72$.]{
    \psfrag{V}{$V_\mathrm{eff}(\phi)$}
    \psfrag{f}{$\phi$}
    \includegraphics[scale=0.2]{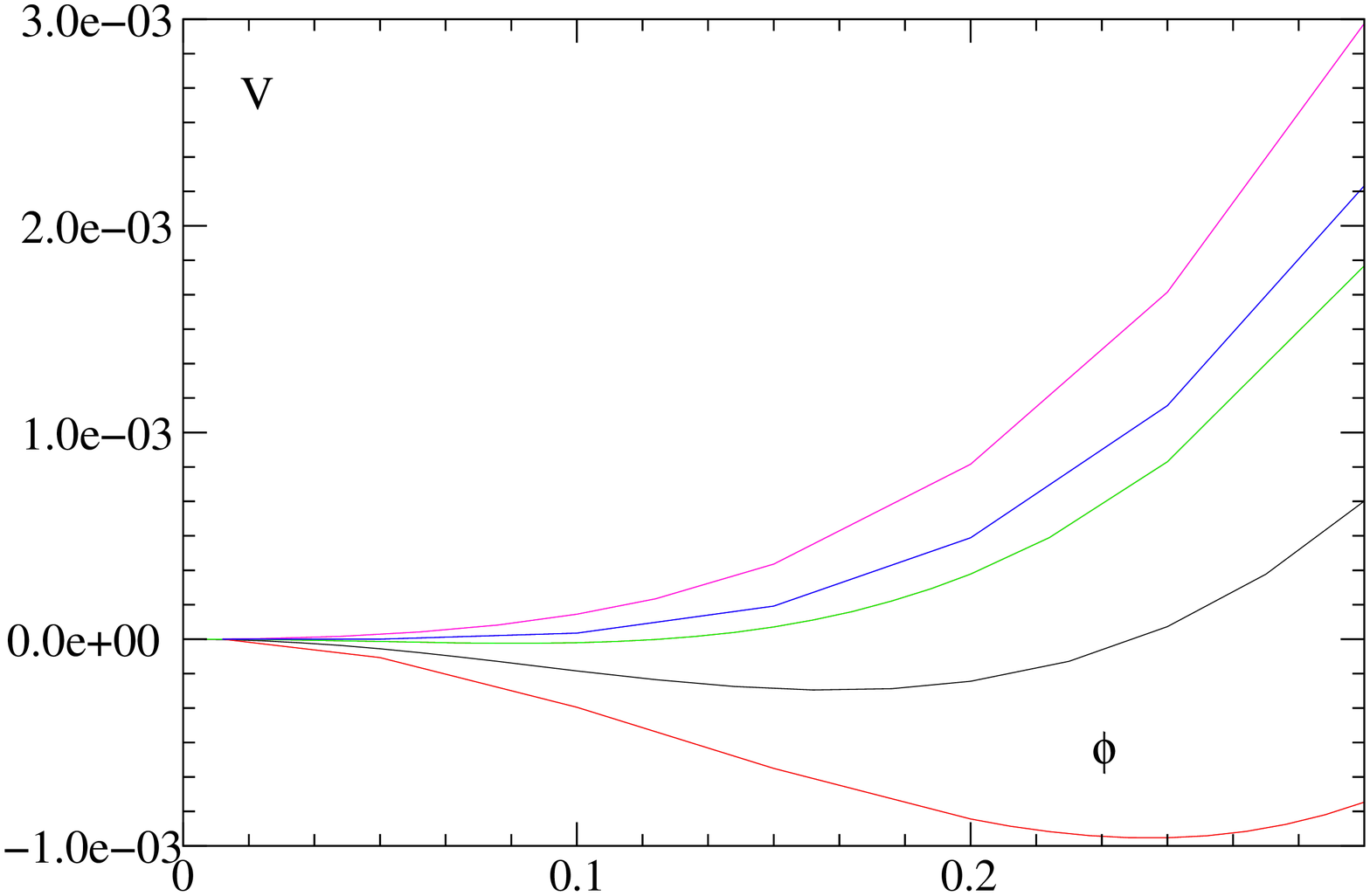}
  }
  \caption{Effective 2PPI potential for $N=4$, $\mu^2=v^2$ and $\lambda=1$.
  The potential clearly shows a first-order phase transition in
  the Hartree approximation whereas it is of second order in the
  2-loop approximation.}
  \label{fig:effpot}
\end{figure}

\begin{figure}[htbp]
  \centering
  \psfragscanon
  \psfrag{v(T)}{$\phi_0(T)$}
  \psfrag{T}{$T$}
  \includegraphics[scale=0.2]{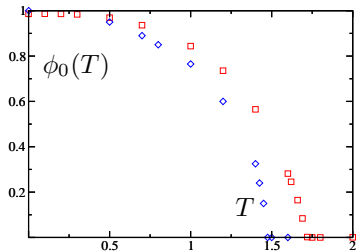}  
  \caption{Temperature-dependence of the minimum $\phi_0(T)$ of the 
    2-loop effective potential. 
    $\lambda=1$ (squares), $\lambda=0.1$ (diamonds).
    The continuous transition to zero is signal for
    a second-order phase transition.}
  \label{fig:vT}
\end{figure}

By inserting the ($\phi$-dependent) solutions of the
gap equations \eqref{eq:gapsigmapi} into Eq.~\eqref{eq:gamma2PPI}
the 1PI effective action is regained. 
The effective potential is then obtained by splitting off an irrelevant
global factor of volume and temperature
  \[ V^\mathrm{1PI}_\mathrm{eff}[\phi] = 
     V^\mathrm{2PPI}_\mathrm{eff}[\phi,\mathcal{M}^2_\sigma(\phi),
     \mathcal{M}^2_\pi(\phi)]
     \ .
  \]
We set $N=4$, fix the renormalization scale to 
$\mu^2=v^2$ and calculate the value of $ V^\mathrm{1PI}_\mathrm{eff}$
for fixed $\phi$ and temperature. In Fig.~\ref{fig:effpot} we
plot the effective potential for different temperatures 
both in the Hartree and in the 2-loop 
approximation. The potential in the Hartree approximation has a
false vacuum -- an evidence for a first-order phase transition --
whereas the 2-loop effective potential shows only one minimum, i.e.,
a phase transition of second order.
We call the (temperature-dependent) minimum of the 2-loop
potential $\phi_0(T)$ and plot its value against temperature 
(cf. Fig.~\ref{fig:vT}). There is a continous transition from
a non-trivial vacuum for temperatures below the critical one 
to a phase with $\phi_0(T)=0$ for temperatures $T>T_\mathrm{crit}$\ .

At first sight this seems to be a contradiction to the result obtained
by Bordag and Skalozub\cite{Bordag} who found
a (weak) first-order phase transition beyond Hartree. However,
they use the 2PI CJT approach to calculate the free energy which 
differs from ours. They obtain a local gap equation from the non-local
Dyson--Schwinger equation by considering it in the zero momentum limit, 
which results in a loss of thermodynamic consistency
because a local propagator does not follow from the 2PI formalism 
by a variational principle. 
It seems that, for some reason, this procedure changes the order of the 
phase transition which cannot be undone, not even by including 
infinitely many diagrams of higher order (see also,e.g., 
Ref.~\cite{Phat:2003eh}).


\section{Variational Mass Parameters and Physical Masses}

\subsection{Variational mass parameters}
\begin{figure}
  \centering
  \psfragscanon
  \psfrag{m}{$\mathcal{M}_{\sigma,\pi}$}
  \psfrag{T}{$T$}
  \psfrag{s,p}{ }
  \includegraphics[scale=0.2]{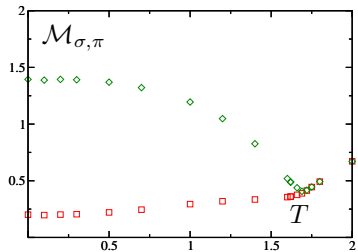}
  \caption{Temperature-dependence of the mass parameters for $\lambda=1$. 
    Diamonds: $\mathcal{M}_\sigma$, squares: $\mathcal{M}_\pi$.}
  \label{fig:MvarT}
  \psfragscanoff
\end{figure}

In Fig.~\ref{fig:MvarT} we show the values of the mass 
parameters $\mathcal{M}_\sigma$ and $\mathcal{M}_\pi$ as 
functions of temperature. Both of them show a smooth
temperature-dependent behavior. They become equal for
temperatures beyond the critical one where the symmetry
is restored. The pion mass parameter is small but never equal to
zero -- even not for zero temperature. Though recalling Goldstone's
theorem this should actually be the case. The violation of this theorem
is due to the fact that -- for practical reasons -- one has to
truncate the effective action at some level and therefore
(slightly) violates symmetries of the underlying theory. For the 
full effective action at the stationary point there should be
an equality between $\delta^2\Gamma / \delta\phi^2$ and
$\mathcal{M}^2$ which is not present here, in the 2-loop approximation,
 as we will show in the following.

\subsection{Physical pion mass}
We will define the \emph{physical} masses as the eigenvalues of the
matrix of second derivatives of the 1PI effective potential.
This physical pion mass is \emph{always} zero 
if the potential is $O(N)$-symmetric. This is obvious when 
the $O(N)$-symmetric physical mass matrix (with the eigenvalues
$M_{\sigma,\mathrm{phys}}^2$ and $M_{\pi,\mathrm{phys}}^2$)
  \[ M^2_{ij,\mathrm{phys}} 
  = \frac{\phi_i\phi_j}{\phi^2} M_{\sigma,\mathrm{phys}}^2
  + \left( \delta_{ij} - \frac{\phi_i\phi_j}{\phi^2} \right) 
  M_{\pi,\mathrm{phys}}^2\]
is compared to the second derivatives of an $O(N)$-symmetric
potential $V(\vec\phi^2)$
  \[ M^2_{ij,\mathrm{phys}} =
  \frac{\partial^2 V(\vec{\phi}^2)}{\partial \phi_i \partial \phi_j}
  \biggr|_{\vec\phi=\vec\phi_0}
  = 2\delta_{ij}\, V'(\vec\phi^2) + 4\phi_i\phi_j\, V''(\vec\phi^2)
  \biggr|_{\vec\phi=\vec\phi_0} \ .\]
Since $\vec\phi_0 = \{\phi_0(T),0,\dots,0\}$ and $M_{\pi,\mathrm{phys}}^2$ 
is obtained by derivatives perpendicular to $\vec\phi_0$ it follows
that
\[  M_{\pi,\mathrm{phys}}^2  = 2\, V'(\vec\phi^2_0) = 0\ .\]
That is how Goldstone's theorem is saved in this context.


\subsection{Physical sigma mass}
\begin{figure}[htbp]
  \centering
  \psfragscanon
  \psfrag{M}{$M_{\sigma,\mathrm{phys}}$}
  \psfrag{s}{}
  \psfrag{T}{$T$}
  \includegraphics[scale=0.2]{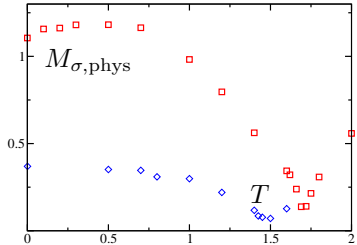}    
  \caption{Physical $\sigma$ mass as a function of temperature.
  $\lambda=1$ (squares), $\lambda=0.1$ (diamonds). }
  \label{fig:sigmamass}
\end{figure}

The physical sigma mass is then given by the second derivative
of the 1PI effective potential
\[M^2_{\sigma,\mathrm{phys}}(T) = \frac{\partial^2 
  V_\mathrm{eff}^\mathrm{1PI}(\phi)}{\partial \phi^2}
  \biggr|_{\phi=\phi_0(T)}\ . \]
We calculated this number numerically fitting the (numerically obtained)
effective potential
with a polynomial. The results are shown in Fig.~\ref{fig:sigmamass}.


\section{Summary and Outlook}
We have analyzed the $O(N)$ linear sigma model in the 2PPI
formalism within the 2-loop approximation, i.e., beyond leading order 
(or Hartree).
As in the $N=1$ version of the model\cite{Smet:2001un} 
the phase transition, which is of first order in the Hartree approximation,
becomes a second-order one. 
Furthermore there is no direct contradiction to the results of Bordag
and Skalozub\cite{Bordag} and Phat~\etal\cite{Phat:2003eh} since they 
use a different method.

As in the Hartree approximation and in
the $N=1$ version we find that the variational mass parameter of the pion
quantum fluctuations is different from zero in the
broken symmetry phase, so that a naive particle interpretation of this
quantity -- suggested by the large-$N$ analysis -- becomes problematic. 
By defining the physical sigma and pion masses as the eigenvalues of
the matrix of second derivatives of the 1PI effective potential,
Goldstone's theorem is saved.

A future step could certainly be to extend this analysis to 
non-equilibrium, i.e.,
to investigate if the system exhibits a second-order phase transition
in the 2-loop approximation out of equilibrium.

\section*{Acknowledgments}
We would like to thank Andreas Heinen and Henri Verschelde
for useful and stimulating 
discussions on the 2PPI formalism.

Furthermore we appreciate several fruitful discussions with
Michael Bordag who gave us interesting insights
into his work.

S.M. thanks the \emph{Graduiertenkolleg ``Physik der Elementarteilchen 
an Be\-schleu\-nigern und im Universum''} for financial support.

\end{document}